\documentclass[letterpaper, 10 pt,conference]{ieeeconf}  
\usepackage{amssymb}
\usepackage{amsmath,mathrsfs}
\usepackage{color}
\usepackage{graphicx}
\usepackage{mathtools}
\usepackage{lineno,hyperref}
\usepackage{cleveref}
\usepackage{bm}
\let\labelindent\relax
\usepackage{enumitem}
\usepackage{tikz}
\usepackage{comment}
\usepackage{algorithm}
\usepackage{algorithmicx}
\usepackage{adjustbox}
\usepackage{algpseudocode}
\modulolinenumbers[5]
\newtheorem{theorem}{Theorem}[section]
\newtheorem{remark}{Remark}[section]
\newtheorem{example}{Example}[section]
\newtheorem{assumption}{Assumption}[section]

\newtheorem{corollary}{Corollary}[section]
\newtheorem{lemma}[theorem]{Lemma}

\usepackage{tikz}

\definecolor{MatlabBlue}{rgb}    {0     , 0.4470, 0.7410}
\definecolor{MatlabRed}{rgb}     {0.8500, 0.3250, 0.0980}
\definecolor{MatlabYellow}{rgb}  {0.9290, 0.6940, 0.1250}
\definecolor{MatlabPurple}{rgb}  {0.4940, 0.1840, 0.5560}
\definecolor{MatlabGreen}{rgb}   {0.4660, 0.6740, 0.1880}
\definecolor{MatlabBabyBlue}{rgb}{0.3010, 0.7450, 0.9330}
\definecolor{MatlabGray}{rgb}{0.5, 0.5, 0.5}
\definecolor{MatlabLightGray}{rgb}{0.75, 0.75, 0.75}
\definecolor{MatlabBlack}{rgb}{0, 0, 0}
\definecolor{MatlabLightGray4}{rgb}{0.875, 0.875, 0.875}
\definecolor{MatlabLightGray3}{rgb}{0.85, 0.85, 0.85}
\definecolor{MatlabLightGray2}{rgb}{0.775, 0.775, 0.775}
\definecolor{MatlabLightGray1}{rgb}{0.7, 0.7, 0.7}
\definecolor{MatlabGray20}{rgb}{0.2, 0.2, 0.2}
\definecolor{MatlabGray30}{rgb}{0.3, 0.3, 0.3}
\definecolor{MatlabGray40}{rgb}{0.4, 0.4, 0.4}
\definecolor{MatlabGray50}{rgb}{0.5, 0.5, 0.5}
\definecolor{MatlabGray60}{rgb}{0.6, 0.6, 0.6}
\definecolor{MatlabGray70}{rgb}{0.7, 0.7, 0.7}
\definecolor{MatlabGray80}{rgb}{0.8, 0.8, 0.8}
\definecolor{MatlabGray85}{rgb}{0.85, 0.85, 0.85}
\definecolor{MatlabGray90}{rgb}{0.9, 0.9, 0.9}

\usepackage{multirow}
\usepackage{color}
\usepackage{cite}

\usepackage{eso-pic}
\AddToShipoutPictureBG*{%
	\AtPageUpperLeft{%
		\setlength\unitlength{1in}%
		\hspace*{\dimexpr0.5\paperwidth\relax}
		\makebox(0,-0.75)[c]{
			\begin{tabular}{c c}
				Rodrigo A. Gonz\'alez \emph{et al.},
				Statistical Analysis of Block Coordinate Descent Algorithms for Linear Continuous-time System Identification. \\
				To appear in
				{\em IEEE Control Systems Letters (L-CSS)},
				2024,
				uploaded to ArXiv on April 13th, 2024 \\
\end{tabular}}}}

\providecommand{\keywords}[1]{\textbf{\textit{Index terms---}}}

\IEEEoverridecommandlockouts                              
\title{\LARGE \bf Statistical Analysis of Block Coordinate Descent Algorithms for Linear Continuous-time System Identification}

\author{Rodrigo A. Gonz\'alez, Koen Classens, Cristian R. Rojas, James S. Welsh, Tom Oomen 
\thanks{This work was supported by the research program VIDI with project number 15698, which is (partly) financed by the Netherlands Organization for Scientific Research (NWO), by the Digital Futures project EXTREMUM, and by the Swedish Research Council under contract number 2016-06079 (NewLEADS). R. A. Gonz\'alez, K. Classens and T. Oomen are with the Department of Mechanical Engineering, Eindhoven University of Technology, Eindhoven, The Netherlands. C. R. Rojas is with the Division of Decision and Control Systems, KTH Royal Institute of Technology, Stockholm, Sweden. J. S. Welsh is with the School of Engineering, University of Newcastle, NSW, Australia. T. Oomen is also with the Delft Center for Systems and Control, Delft University of Technology, Delft, The Netherlands. E-mail of the corresponding author: r.a.gonzalez@tue.nl.
}%
}
\begin{document}

\maketitle

\begin{abstract}
Block coordinate descent is an optimization technique that is used for estimating multi-input single-output (MISO) continuous-time models, as well as single-input single output (SISO) models in additive form. Despite its widespread use in various optimization contexts, the statistical properties of block coordinate descent in continuous-time system identification have not been covered in the literature. The aim of this paper is to formally analyze the bias properties of the block coordinate descent approach for the identification of MISO and additive SISO systems. We characterize the asymptotic bias at each iteration, and provide sufficient conditions for the consistency of the estimator for each identification setting. The theoretical results are supported by simulation examples.
\end{abstract}
\vspace{-0.17cm}
\section{Introduction}
\label{sec:introduction}
System identification deals with the problem of learning dynamical models using input and output data \cite{ljung1998system}. In this field, a fundamental choice is whether to work with discrete-time or continuous-time models based on the specific application. Continuous-time system identification \cite{Garnier2008book} is advantageous if a requirement is that the model should have physical significance, and its methods can be adjusted to address irregular sampling and the estimation of stiff models~\cite{garnier2014advantages}.

For identifying continuous-time, single-input single-output (SISO) systems, a widely-used method is the simplified refined instrumental variable method (SRIVC, \cite{young1980refined}), which iteratively computes instrumental variables estimates of the system parameters. A limitation is that the pseudolinear equation used in the SRIVC estimator cannot handle multiple-input single-output (MISO) models described with transfer functions of distinct denominator polynomials. For such case, \cite{garnier2007optimal} (see also \cite[Sec. 5.2]{young1980refined}) proposes a backfitting algorithm that estimates each submodel keeping all the other submodels fixed, and iterating until convergence. Such algorithm has been applied to model blood glucose \cite{kirchsteiger2011direct} and wireless power systems \cite{chen2020data}, among other applications. Recently, a similar identification method for additive SISO models was introduced in \cite{gonzalez2023parsimonious}. This additive representation is particularly useful for systems interpreted in modal form, common in many mechanical systems. Unlike the approach presented in \cite{garnier2007optimal}, in \cite{gonzalez2023parsimonious}, instrumental variable iterations are performed for each submodel until convergence, aligning with the philosophy of the block coordinate descent approach for non-convex optimization~\cite{luenberger2008linear}. This descent strategy has been also used for optimization over networks \cite{necoara2017random}, and for training Neural ODEs \cite{quaglino2020snode}.

The statistical analysis of continuous-time system identification methods is essential for their applicability, with substantial progress being made in recent years. The consistency and efficiency of the SRIVC estimator are studied in \cite{pan2020consistency} and \cite{pan2020efficiency}, and its relation to Maximum Likelihood is studied in \cite{young2015refined}. Likewise, \cite{gonzalez2022consistency} extended these findings to closed-loop models. Despite this progress, the block coordinate descent method for MISO and additive SISO identification does not yet have this theoretical foundation. The method for MISO systems is believed to yield consistent estimates upon convergence \cite[Section 3.3]{garnier2007optimal}, though it lacks a formal proof. Similarly, conditions for consistent estimates have not been pursued in \cite{gonzalez2023parsimonious}. Although Monte Carlo simulations can provide evidence of the performance of these methods, a comprehensive theoretical analysis of their bias properties is essential to offer practical guidelines for their use under general conditions, and may help to design identification experiments that allow for consistent estimators.

In this paper, we analyze the statistical properties of the backfitting methods proposed in \cite{garnier2007optimal} and \cite{gonzalez2023parsimonious} in a unified manner. Our main contributions are outlined below.
\vspace{-0.03cm}
\begin{enumerate}[label=C\arabic*]
	\item
	\label{contributionC1}
    We generalize the algorithms in \cite{garnier2007optimal} and \cite{gonzalez2023parsimonious} by deriving closed-form expressions for the Gauss-Newton (GN) and SRIVC iterations, which, upon convergence for finite sample size, are proven to deliver a stationary point of the cost function under mild conditions. 
    \item 
    \label{contributionC2}
    We characterize the asymptotic bias obtained at each descent iteration, which shows that consistent estimates can be achieved in one iteration in the MISO setup case.
    \item 
    \label{contributionC3}
    We discuss the difference between the MISO and additive SISO setups in terms of the persistence of excitation conditions required for consistent estimates, and we provide sufficient conditions for the generic consistency of the block coordinate descent approach for each setup.
\end{enumerate}
\vspace{-0.03cm}
The remainder of this paper is structured as follows. Section \ref{sec:system} introduces the identification setup. Section \ref{sec:bcd} revisits the block coordinate descent method that is used for identifying MISO and additive SISO models. The statistical analysis is addressed in Section \ref{sec:open-loop}. Extensive simulations that verify the theoretical results can be found in Section \ref{sec:simulations}, while Section \ref{sec:conclusions} contains concluding remarks. 

\textit{Notation}: Vectors and matrices are written in bold. The Heaviside operator $p$ is defined as $pu(t) = \frac{\textnormal{d}}{\textnormal{d}t}u(t)$, and $G(p)u(t_k)$ means that the continuous-time signal $u(t)$ is filtered through the transfer function $G(p)$ and evaluated at $t=t_k$. If $A(p)$ and $B(p)$ are polynomials of degrees $n$ and $m$ respectively, then $\mathcal{S}(A,B)$ is a $(n+m+1)\times (n+m+1)$ Sylvester matrix (see, e.g., Eq. (A3.9) of \cite{soderstrom1983instrumental}).

\section{Additive SISO and MISO system setups}
\label{sec:system}
Consider the following $K\in\mathbb{N}$ linear and time-invariant, causal, continuous-time systems
\begin{equation}
    G_i^*(p) = \frac{B^*_i(p)}{A_i^*(p)}, \quad i=1,\dots,K.
\end{equation}
The numerator and denominator polynomials $B_i^*(p)$ and $A_i^*(p)$ are assumed to be coprime, i.e., they do not share roots, and
\begin{equation}
	\label{parametrization}
\begin{split}
	A_i^*(p)\hspace{-0.05cm}&= \hspace{-0.03cm}a_{i,n_i}^*p^{n_i}\hspace{-0.05cm}+\hspace{-0.05cm}a_{i,n_i-1}^*p^{n_i-1}\hspace{-0.05cm}+\cdots + a_{i,1}^*p +\hspace{-0.03cm} 1,  \hspace{-0.1cm}\\
	B_i^*(p)\hspace{-0.05cm}&=\hspace{-0.03cm} b_{i,m_i}^*p^{m_i}\hspace{-0.05cm}+\hspace{-0.05cm}b_{i,m_i-1}^*p^{m_i-1}\hspace{-0.05cm}+\cdots + b_{i,1}^*p \hspace{-0.03cm}+ \hspace{-0.03cm}b_{i,0}^*,	\hspace{-0.1cm}
\end{split}
\end{equation}
with $a_{i,n_i}^*\neq 0$ and $m_i\leq n_i$. The polynomials $A_i^*(p)$ and $B_i^*(p)$ are jointly described by the parameter vector 
\begin{equation}
	\label{ctparametervector}
	\bm{\theta}_i^* = \begin{bmatrix}
		a_{i,1}^*, & \hspace{-0.2cm}a_{i,2}^*, & \hspace{-0.2cm} \dots, & \hspace{-0.2cm}a_{i,n_i}^*, & \hspace{-0.2cm}b_{i,0}^*, & \hspace{-0.2cm}b_{i,1}^*, & \hspace{-0.2cm}\dots, & \hspace{-0.2cm}b_{i,m_i}^*
	\end{bmatrix}^\top.
\end{equation}
These systems are excited by inputs $u_i(t)$, $i=1,\dots,K$, and a noisy measurement of the sum of the outputs is retrieved at every time instant $t=t_k$, $k=1,\dots,N$, where $\{t_k\}_{k=1}^N$ are assumed to be evenly spaced in time. That is,
\begin{align}
	\label{system1}
	x(t) &= \textstyle\sum_{i=1}^K G_i^*(p) u_i(t)  \\
	y(t_k) &= x(t_k) + v(t_k),
\end{align}
where $v(t_k)$ is a zero-mean stationary random process of finite variance. Two frameworks are studied in this paper: 
\begin{enumerate}
    \item Continuous-time MISO identification: We assume that the inputs $u_i$ are uncorrelated with each other, and we allow the denominator polynomials $A_i$ to be distinct.
    \item Continuous-time additive SISO identification: We assume that all inputs $u_i$ are equal. In addition, 
    we assume that at most one subsystem is biproper.
\end{enumerate}
This paper presents a statistical analysis of block coordinate algorithms to obtain estimates of the joint parameter vector of the systems $G_i^*(p)$, i.e.,
\begin{equation}
	\bm{\beta}^*:= \begin{bmatrix}
		\bm{\theta}_1^{*\top}, & \hspace{-0.2cm}\bm{\theta}_2^{*\top}, & \hspace{-0.2cm}\dots, & \hspace{-0.2cm} \bm{\theta}_K^{*\top}
	\end{bmatrix}^\top. \notag
\end{equation}
We consider $\{u_1(t_k),\dots,u_K(t_k),y(t_k)\}_{k=1}^N$ as given. In the next section, we revisit the block coordinate descent algorithm that solves this identification problem for both MISO and additive SISO frameworks.

\section{Block coordinate descent approach for continuous-time system identification}
\label{sec:bcd}

Towards the goal of computing an estimator for $\bm{\beta}^*$, we are interested in solving the following minimization problem:
\begin{align}
	\label{cost}
	\min_{\substack{\bm{\theta}_i \in \Omega_i,\\ i = 1, \dots, K}} V(\bm{\beta}),
\end{align}
where $\Omega_i\subset \mathbb{R}^{n_i+m_i+1}$ is a compact set where the true parameters of the $i$th subsystem are assumed to lie, and 
\begin{equation}
	\label{l2openloop}
	V (\bm{\beta}) = \frac{1}{N}\sum_{k=1}^N \bigg[y(t_k) -\sum_{i=1}^K G_i(p,\bm{\theta}_i) u_i(t_k) \bigg]^{2}.
\end{equation}
The approach for solving \eqref{cost} that we investigate consists of iteratively minimizing the cost with respect to $\bm{\theta}_i$ while leaving the other variables fixed. Algorithm \ref{algorithm1} describes the general procedure, known as \emph{block coordinate descent} \cite{luenberger2008linear}.
\vspace{-0.25cm}
\begin{algorithm}
	\caption{Block coordinate descent}
	\begin{algorithmic}[1]
		\State Choose an initial parameter vector $\bm{\theta}_i^1$ for each $i$
		\For{$l = 1, 2, \dots$}
		\For{$i = 1, \dots, K$}
		\State $\hspace{-0.3cm}\bm{\theta}_i^{l\hspace{-0.01cm}+\hspace{-0.01cm}1} \hspace{-0.12cm}\gets \hspace{-0.06cm} \arg\hspace{-0.02cm}\min\limits_{\hspace{-0.53cm}\bm{\theta}_i \in \Omega_i} \hspace{-0.03cm} V_{\hspace{-0.02cm}N}\hspace{-0.02cm}(\bm{\theta}_1^{l+1}\hspace{-0.02cm},\hspace{-0.02cm} \dots\hspace{-0.02cm}, \bm{\theta}_{i-1}^{l+1}\hspace{-0.02cm},\hspace{-0.03cm} \bm{\theta}_i\hspace{-0.02cm},\hspace{-0.02cm} \bm{\theta}_{i+1}^{l}\hspace{-0.02cm},\hspace{-0.02cm} \dots\hspace{-0.02cm},\hspace{-0.02cm} \bm{\theta}_K^{l}\hspace{-0.02cm})$
		\EndFor
		\EndFor
	\end{algorithmic}
	\label{algorithm1}
\end{algorithm}

\vspace{-0.3cm}
For each iteration (in $l$) of Algorithm \ref{algorithm1}, denote $\bm{\beta}^l := [(\bm{\theta}_1^l)^\top, \dots, (\bm{\theta}_K^l)^\top]^\top$ and the update rule $\bm{\beta}^{l+1} = \mathcal{A}(\bm{\beta}^l)$. The following result concerns the convergence of Algorithm \ref{algorithm1} to a stationary point of the cost \eqref{cost}.

\begin{theorem}[Global convergence of Algorithm \ref{algorithm1}]
	\label{thm42}
	If the search along any coordinate direction $\bm{\theta}_i$ yields a unique minimum point of $V$, then the limit of any convergent subsequence of $\{\bm{\beta}^l\}$ obtained from $\bm{\beta}^{l+1} = \mathcal{A}(\bm{\beta}^l)$ belongs to the set of fixed points $\{\bm{\beta}\in \hspace{-0.05cm}\prod_{i=1}^K\Omega_i \colon \nabla V(\bm{\beta}) \hspace{-0.05cm}=\hspace{-0.05cm} \mathbf{0}\}$.
\end{theorem}
\begin{proof}
See, e.g., \cite[Section 8.9]{luenberger2008linear}.     
\end{proof}

\begin{remark}
	\label{remarkglobal}
    The result in Theorem \ref{thm42} still holds if the unique minimum assumption is replaced with the requirement that $V$ decreases at each coordinate search.
\end{remark}

\vspace{-0.1cm}

\section{Statistical Analysis}
\label{sec:open-loop}
\vspace{-0.1cm}
This section addresses the statistical properties of Algorithm \ref{algorithm1} applied to the identification of MISO and additive SISO continuous-time systems. We explore Gauss-Newton and refined instrumental variable methods for computing the descent, the parameter bias at each descent iteration, persistence of excitation issues, and statistical consistency. These topics will be covered in the following four subsections.

\vspace{-0.15cm}
\subsection{Two alternatives for computing the descent step}
\vspace{-0.05cm}
The block coordinate descent algorithm described in Algorithm \ref{algorithm1} requires a way to compute, at each iteration,
\vspace{-0.05cm}
\begin{align}
	\bm{\theta}_i^{l+1} &= \underset{\bm{\theta}_i \in \Omega_i}{\arg \min} \frac{1}{N} \sum_{k=1}^N \bigg[ y(t_k) - \sum_{j=1}^{i-1} G_j(p,\bm{\theta}_j^{l+1}) u_j(t_k) \notag \\
	\label{opt}
	&-\hspace{-0.12cm}\sum_{j=i+1}^{K} \hspace{-0.05cm}G_j(p,\bm{\theta}_j^l) u_j(t_k) - G_i(p,\bm{\theta}_i) u_i(t_k) \bigg]^2
\end{align}
for $i=1,2,\dots, K$. For fixed values of $\{\bm{\theta}_j^{l+1}\}_{j=1}^{i-1}$ and $\{\bm{\theta}_j^l\}_{j=i+1}^{K}$,  and for a sufficiently large compact parameter space $\Omega_i$, the optimization problem in \eqref{opt} reduces to a nonlinear least-squares problem that can be solved via iterations. Indeed, if we define the residual output of each submodel
\begin{equation}
	\tilde{y}_i(t_k) \hspace{-0.09cm}:=\hspace{-0.03cm} y(t_k)\hspace{-0.01cm}-\hspace{-0.02cm} \sum_{j=1}^{i-1} \hspace{-0.06cm}G_{\hspace{-0.02cm}j}\hspace{-0.03cm}(p,\hspace{-0.02cm}\bm{\theta}_j^{l\hspace{-0.02cm}+\hspace{-0.02cm}1}\hspace{-0.02cm}) u_j(t_k)\hspace{-0.01cm}-\hspace{-0.1cm}\sum_{j=i+1}^{K} \hspace{-0.15cm}G_{\hspace{-0.02cm}j}\hspace{-0.02cm}(p,\hspace{-0.02cm}\bm{\theta}_j^l) u_j(t_k), \notag
\end{equation}
then $\bm{\theta}_i^{l+1}$ must satisfy the first-order optimality condition
\begin{equation}
	\label{convergingto}
	\frac{1}{N}\sum_{k=1}^N \hat{\bm{\varphi}}_\textnormal{f}(t_k,\bm{\theta}_i^{l+1}) e(t_k,\bm{\theta}_i^{l+1}) = \mathbf{0},
\end{equation}
where the gradient and total residual are denoted as
\begin{align}
	&\hspace{-0.23cm}\hat{\bm{\varphi}}_\textnormal{f}(\hspace{-0.01cm}t_k,\hspace{-0.01cm}\bm{\theta}_i^{l\hspace{-0.01cm}+\hspace{-0.01cm}1}\hspace{-0.01cm}) \hspace{-0.11cm}=\hspace{-0.13cm} \bigg[\hspace{-0.02cm}\frac{-\hspace{-0.01cm}p\hspace{-0.01cm}B_i^{l\hspace{-0.01cm}+\hspace{-0.01cm}1}(p)}{[A_i^{l\hspace{-0.01cm}+\hspace{-0.01cm}1}(p)]^2}\hspace{-0.02cm} u_i(\hspace{-0.01cm}t_k\hspace{-0.01cm})\hspace{-0.01cm}, \hspace{-0.02cm}\dots\hspace{-0.02cm}, \hspace{-0.08cm}\frac{-\hspace{-0.01cm}p^{\hspace{-0.01cm}n_i} \hspace{-0.05cm}B_i^{l\hspace{-0.02cm}+\hspace{-0.02cm}1}(p)}{[A_i^{l\hspace{-0.02cm}+\hspace{-0.02cm}1}(p)]^2} u_i(\hspace{-0.01cm}t_k\hspace{-0.01cm}), \notag \\
	\label{filteredinstrument}
	&\hspace{1.7cm} \frac{1}{A_i^{l\hspace{-0.02cm}+\hspace{-0.02cm}1}(p)}u_i(t_k),\dots, \hspace{-0.03cm} \frac{p^{m_i}}{A_i^{l\hspace{-0.02cm}+\hspace{-0.02cm}1}(p)} u_i(t_k) \hspace{-0.02cm}\bigg]^{\hspace{-0.07cm}\top}\hspace{-0.07cm}, \hspace{-0.1cm} \\
	\label{residual}
	&\hspace{-0.1cm}e(t_k,\bm{\theta}_i^{l\hspace{-0.02cm}+\hspace{-0.02cm}1}) \hspace{-0.06cm}=\hspace{-0.04cm} \tilde{y}_i(t_k)- \frac{B_i^{l\hspace{-0.02cm}+\hspace{-0.02cm}1}(p)}{A_i^{l\hspace{-0.02cm}+\hspace{-0.02cm}1}(p)}u_i(t_k),
\end{align}
with $B_i^{l+1},A_i^{l+1}$ denoting, respectively, the numerator and denominator polynomials of the $i$th model evaluated at $\bm{\theta}_{i}^{l+1}$. Lemma \ref{lemmagnsrivc} provides two alternatives that are proven to deliver the \textit{same} stationary points of the cost function in \eqref{opt} at convergence in iterations under mild conditions.
\begin{lemma}[Gauss-Newton and SRIVC iterations]
	\label{lemmagnsrivc}
	For an initial model parameter estimate $\bm{\theta}_{i,0}^{l+1}$ and $s=0,1,2,\dots$, consider the Gauss-Newton (GN) iterations
		\begin{align}
		\label{gnformula}
			\bm{\theta}_{i,s+1}^{l+1} &= \bigg[\frac{1}{N}\sum_{k=1}^{N}\hat{\bm{\varphi}}_\textnormal{f}(t_k,\bm{\theta}_{i,s}^{l+1}) \hat{\bm{\varphi}}_\textnormal{f}^\top(t_k,\bm{\theta}_{i,s}^{l+1})  \bigg]^{-1} \\
			&\hspace{-0.75cm}\times \hspace{-0.06cm} \left[\hspace{-0.03cm}\frac{1}{N}\hspace{-0.03cm}\sum_{k=1}^{N}\hspace{-0.03cm}\hat{\bm{\varphi}}_\textnormal{f}(t_k,\bm{\theta}_{i,s}^{l+1}) \hspace{-0.02cm}\bigg(\hspace{-0.04cm}e(t_k,\hspace{-0.01cm}\bm{\theta}_{i,s}^{l+1})\hspace{-0.045cm}+\hspace{-0.045cm}\frac{B_{i,s}^{l+1}(p)}{[A_{i,s}^{l+1}(p)]^2}u_i(t_k)\hspace{-0.03cm}\bigg) \hspace{-0.02cm} \right]\hspace{-0.05cm}, \notag
		\end{align}
	and the SRIVC \cite{young1980refined} iterations
		\begin{align}
			\bm{\theta}_{i,s+1}^{l+1} &= \bigg[\frac{1}{N}\sum_{k=1}^{N}\hat{\bm{\varphi}}_\textnormal{f}(t_k,\bm{\theta}_{i,s}^{l+1}) \bm{\varphi}_\textnormal{f}^\top(t_k,\bm{\theta}_{i,s}^{l+1})\bigg]^{-1} \notag \\
			\label{formulasrivc}
            &\hspace{0.5cm}\times \hspace{-0.04cm} \bigg[\frac{1}{N}\sum_{k=1}^{N}\hat{\bm{\varphi}}_\textnormal{f}(t_k,\bm{\theta}_{i,s}^{l+1}) \tilde{y}_{\textnormal{f}}(t_k,\bm{\theta}_{i,s}^{l+1})\bigg],
		\end{align}
	where $\bm{\varphi}_\textnormal{f}$ and filtered residual output $\tilde{y}_{\textnormal{f}}$ are given by
	\begin{align}
		&\hspace{-0.2cm}\bm{\varphi}_\textnormal{f}(t_k,\bm{\theta}_{i,s}^{l\hspace{-0.01cm}+\hspace{-0.01cm}1})=  \bigg[\frac{-p}{A_{i,s}^{l\hspace{-0.01cm}+\hspace{-0.01cm}1}(p)} \tilde{y}_i(t_k), \dots, \hspace{-0.01cm}\frac{-p^{n_i}}{A_{i,s}^{l\hspace{-0.01cm}+\hspace{-0.01cm}1}(p)} \tilde{y}_i(t_k), \notag \\
		\label{filteredregressor}
		&\hspace{1.6cm} \frac{1}{A_{i,s}^{l\hspace{-0.01cm}+\hspace{-0.01cm}1}(p)}u_i(t_k),\dots, \hspace{-0.01cm} \frac{p^{m_i}}{A_{i,s}^{l\hspace{-0.01cm}+\hspace{-0.01cm}1}(p)} u_i(t_k) \hspace{-0.01cm}\bigg]^{\hspace{-0.01cm}\top}\hspace{-0.05cm}, \hspace{-0.1cm} \\
		\label{filteredoutput}
		&\hspace{-0.2cm}\tilde{y}_\textnormal{f}(t_k,\bm{\theta}_{i,s}^{l+1}) = \frac{1}{A_{i,s}^{l+1}(p)}\tilde{y}_i(t_k).
	\end{align}
	If the matrices being inverted in \eqref{gnformula} and \eqref{formulasrivc} are non-singular for all large enough integers $s$, then any converging point (when $s$ tends to infinity) satisfies \eqref{convergingto}.
\end{lemma}
\begin{proof}
We omit the $(\cdot)_i$ notation on $\bm{\theta}, A,B,$ and $u$ for simplicity, assuming that the expressions refer to the $i$th submodel. Upon convergence as $s\to \infty$, any converging point $\bm{\theta}^{l+1}:=\lim_{s\to \infty} \bm{\theta}_{s}^{l+1}$ of the GN iterations satisfies
\vspace{-0.1cm}
\begin{align}
	\bm{\theta}^{l+1} &= \bigg[\frac{1}{N}\sum_{k=1}^{N}\hat{\bm{\varphi}}_\textnormal{f}(t_k,\bm{\theta}^{l+1}) \hat{\bm{\varphi}}_\textnormal{f}^\top(t_k,\bm{\theta}^{l+1})  \bigg]^{-1} \notag \\
	&\hspace{-0.6cm}\times \hspace{-0.08cm} \bigg[\hspace{-0.02cm}\frac{1}{N}\hspace{-0.04cm}\sum_{k=1}^{N}\hspace{-0.04cm}\hat{\bm{\varphi}}_\textnormal{f}(t_k,\bm{\theta}^{l+1}) \bigg(e(t_k,\bm{\theta}^{l+1})\hspace{-0.05cm}+\hspace{-0.07cm}\frac{B^{l+1}(p)}{[A^{l+1}(p)]^2}u(t_k)\hspace{-0.02cm}\bigg)\hspace{-0.03cm}\bigg], \notag
\end{align}
which, thanks to the non-singularity assumption of the normal matrix above, is equivalent to
\vspace{-0.2cm}
\begin{align}
	\frac{1}{N}\sum_{k=1}^{N}\hat{\bm{\varphi}}_\textnormal{f}(t_k,\bm{\theta}^{l+1}) &\bigg(\hat{\bm{\varphi}}_\textnormal{f}^\top(t_k,\bm{\theta}^{l+1})\bm{\theta}^{l+1} \notag \\
	\label{conditionlemma2}
	&\hspace{-1.5cm}-e(t_k,\bm{\theta}^{l+1})-\frac{B^{l\hspace{-0.02cm}+\hspace{-0.02cm}1}(p)}{[A^{l\hspace{-0.02cm}+\hspace{-0.02cm}1}(p)]^2}u(t_k)\bigg) = \mathbf{0}. 
\end{align}
The gradient vector $\hat{\bm{\varphi}}_\textnormal{f}(t_k,\bm{\theta}^{l+1})$ can be shown to satisfy $\hat{\bm{\varphi}}_\textnormal{f}^{\top}\hspace{-0.06cm}(t_k,\hspace{-0.02cm}\bm{\theta}^{l\hspace{-0.01cm}+\hspace{-0.01cm}1}\hspace{-0.01cm})\bm{\theta}^{l\hspace{-0.01cm}+\hspace{-0.01cm}1}\hspace{-0.1cm} =\frac{B^{l+1}}(p){[A^{l+1}(p)]^2}u(t_k)$. Replacing this expression in \eqref{conditionlemma2} leads to \eqref{convergingto}, which is what we aimed to prove. For the SRIVC iterations, the expression \eqref{formulasrivc} reduces to the following condition when $\bm{\theta}_{s}^{l+1}=\bm{\theta}_{s+1}^{l+1}=\bm{\theta}^{l+1}$:
\vspace{-0.3cm}
\begin{equation}
	\frac{1}{N}\hspace{-0.04cm}\sum_{k=1}^N\hspace{-0.04cm}\hat{\bm{\varphi}}_\textnormal{f}(t_k,\bm{\theta}^{l+1}) \big[\hspace{0.02cm}\tilde{y}_\textnormal{f}(t_k,\bm{\theta}^{l+1})\hspace{-0.04cm}-\hspace{-0.04cm}\bm{\varphi}_\textnormal{f}^{\top}\hspace{-0.04cm}(t_k,\bm{\theta}^{l+1})\bm{\theta}^{l+1}\big] \hspace{-0.04cm}= \mathbf{0}. \notag
\end{equation}
However, from the definitions of $\bm{\varphi}_\textnormal{f}$ and $\tilde{y}_\textnormal{f}$ in \eqref{filteredregressor} and \eqref{filteredoutput},
\vspace{-0.2cm}
\begin{equation}
	\label{et_k}
    \tilde{y}_\textnormal{f}(t_k,\bm{\theta}^{l+1})-\bm{\varphi}_\textnormal{f}^{\top}(t_k,\bm{\theta}^{l+1})\bm{\theta}^{l+1}=e(t_k,\bm{\theta}^{l+1}).
\end{equation}
Substituting this result above also leads to \eqref{convergingto}.    
\end{proof}

In the literature of both the MISO \cite{garnier2007optimal} and additive SISO \cite{gonzalez2023parsimonious} methods, only SRIVC iterations have been considered. Lemma \ref{lemmagnsrivc} derives the Gauss-Newton iterations that also lead to the same stationary points as the SRIVC iterations. While the SRIVC iterations have been known to be numerically robust against poor initialization \cite{Garnier2008book}, the iterations depend on the specification of the intersample behavior of the output, which is not known beforehand, since only output samples are measured. The output intersample behavior can influence the convergence rate of the iterations, but it does not affect the converging point of the algorithm \cite{pan2020consistency}.
\vspace{-0.1cm}
\subsection{Bias analysis at a fixed descent iteration}
For the following analysis, we assume that we compute the GN \eqref{gnformula} or  SRIVC iterations \eqref{formulasrivc} until convergence, and study the converging point of the $i$th submodel for a \textit{fixed} descent iteration $l$ and large sample size. The theoretical analysis requires the following assumptions:
\begin{assumption}[Persistence of excitation]
	\label{assumption51}
	Each input $u_i(t_k)$ is persistently exciting of order no less than $2n_i$.
\end{assumption}
\begin{assumption}[Stationarity]
	\label{assumption515}
	The inputs $u_i(t_k)$ and disturbance $v(t_s)$ are stationary, and each input is independent of the disturbance for all integers $k$ and $s$.
\end{assumption}
\begin{assumption}[Stability and coprimeness]
	\label{assumption52}
	For every integer $s$, the model estimate $B_{i,s}^{l+1}(p)/A_{i,s}^{l+1}(p)$ is stable, and the numerator and denominator polynomials are coprime.
\end{assumption}
\begin{assumption}[Non-singularity]
	\label{assumption53}
	The normal matrix of the GN iterations, $\mathbb{E}\{\hat{\bm{\varphi}}_\textnormal{f}(t_k,\bm{\theta}_{i}^{l+1}) \hat{\bm{\varphi}}_\textnormal{f}^{\top}(t_k,\bm{\theta}_{i}^{l+1})\}$, or the modified normal matrix of the SRIVC iterations, $\mathbb{E}\{\hat{\bm{\varphi}}_\textnormal{f}(t_k,\bm{\theta}_{i}^{l+1}) \bm{\varphi}_\textnormal{f}^{\top}(t_k,\bm{\theta}_{i}^{l+1})\}$, are generically non-singular with respect to all system and model denominators.
\end{assumption}

Assumptions \ref{assumption51} to \ref{assumption52} are standard in the literature on statistical properties of the SRIVC method for open and closed loop \cite{pan2020consistency,gonzalez2022consistency}, and extensive practical experience indicates that instabilities rarely arise, but they may occur if the data is poor \cite{young2012recursive}. Assumption \ref{assumption53} for the SRIVC iterations can be posed as a 2-norm condition on the interpolation error of the output intersample behavior \cite{pan2020consistency}. Since this condition holds for most cases of interest, Assumption \ref{assumption53} is not restrictive.

\begin{theorem}
	\label{thmmiso}
	Consider the block coordinate descent method in Algorithm \ref{algorithm1}, where $V$ is computed using \eqref{l2openloop}, and the coordinate minimization is performed using the GN or SRIVC iterations in Lemma \ref{lemmagnsrivc}. Suppose Assumptions \ref{assumption51} to \ref{assumption53} hold, and define $\bm{\eta}_j\in \mathbb{R}^{n_j+m_j+1}$ as the vector that contains the coefficients of $A_j^*\hspace{-0.02cm}(p)\bar{B}_j(p)-\bar{A}_j(p)B_j^*\hspace{-0.02cm}(p)$ in descending order of degree, where
	\begin{equation}
		\label{notationAjBj}
		\hspace{-0.22cm}[\bar{A}_j\hspace{-0.02cm}(p),\hspace{-0.02cm}\bar{B}_{j}\hspace{-0.02cm}(p)] \hspace{-0.07cm}=\hspace{-0.09cm} \begin{cases}
			\hspace{-0.03cm}[A_{j}^{l\hspace{-0.02cm}+\hspace{-0.02cm}1}\hspace{-0.02cm}(p),B_{j}^{l\hspace{-0.02cm}+\hspace{-0.02cm}1}\hspace{-0.02cm}(p)], & \hspace{-0.16cm} j \hspace{-0.04cm}=\hspace{-0.04cm} 1,\dots, i \hspace{-0.05cm} \\
			\hspace{-0.03cm}[A_{j}^{l}(p),B_{j}^{l}(p)], & \hspace{-0.16cm}j \hspace{-0.04cm}=\hspace{-0.04cm} i\hspace{-0.04cm}+\hspace{-0.04cm}1,\dots, K. \hspace{-0.05cm}
		\end{cases}
	\end{equation}
	If the GN or SRIVC iterations of the $i$th model converge (for a fixed descent iteration $l$) for all $N$ sufficiently large, then the converging point of the $i$th submodel satisfies
	\begin{equation}
		\label{conditionadditive}
		\sum_{j=1}^K \mathbb{E}\bigg\{\frac{1}{\bar{A}_i^2(p)}\mathbf{u}_i(t_k)\frac{1}{A_j^*(p)\bar{A}_j(p)}\mathbf{u}_j^\top(t_k) \bigg\}\bm{\eta}_j = \mathbf{0},
	\end{equation}
	where $\mathbf{u}_j(t_k)$ ($j=1,\dots,K$) is given by
	\begin{equation}
		\label{ujt_k}
		\mathbf{u}_j(t_k) = \begin{bmatrix}
			p^{n_j+m_j}, & p^{n_j+m_j-1}, & \dots, &  1 
		\end{bmatrix}^\top u_j(t_k). 
	\end{equation}
\end{theorem}
\begin{proof}
Define $\bm{\theta}_{i}^{l+1}:= \lim_{s\to \infty}\bm{\theta}_{i,s}^{l+1}$, where $\bm{\theta}_{i,s}^{l+1}$ is computed from the GN \eqref{gnformula} or SRIVC \eqref{formulasrivc} iterations. 
For a fixed sample size $N$, the converging point $\bm{\theta}_{i}^{l+1}$ must satisfy \eqref{conditionlemma2} in the case of the GN iterations, or
	\begin{align}
		\label{finiteall}
		&\bigg[\frac{1}{N}\sum_{k=1}^{N}\hat{\bm{\varphi}}_\textnormal{f}(t_k,\bm{\theta}_{i}^{l+1}) \bm{\varphi}_\textnormal{f}^\top(t_k,\bm{\theta}_{i}^{l+1})\bigg]^{-1} \\
		&\hspace{0.03cm}\times \hspace{-0.1cm}\bigg[ \hspace{-0.05cm}\frac{1}{N} \hspace{-0.07cm}\sum_{k=1}^{N}\hspace{-0.06cm}\hat{\bm{\varphi}}_\textnormal{f}(\hspace{-0.01cm}t_k,\hspace{-0.02cm}\bm{\theta}_{i}^{l\hspace{-0.02cm}+\hspace{-0.02cm}1}\hspace{-0.02cm}) \hspace{-0.06cm}\left(\hspace{-0.02cm}\tilde{y}_{\textnormal{f}}(\hspace{-0.01cm}t_k,\hspace{-0.02cm}\bm{\theta}_{i}^{l\hspace{-0.02cm}+\hspace{-0.02cm}1}\hspace{-0.02cm})\hspace{-0.07cm}-\hspace{-0.06cm}\bm{\varphi}_\textnormal{f}^{\hspace{-0.06cm}\top}\hspace{-0.05cm}(\hspace{-0.01cm}t_k,\hspace{-0.02cm}\bm{\theta}_{i,s}^{l\hspace{-0.02cm}+\hspace{-0.02cm}1}\hspace{-0.02cm})\bm{\theta}_{i}^{l\hspace{-0.02cm}+\hspace{-0.02cm}1}\hspace{-0.02cm}\right)\hspace{-0.06cm}\bigg]\hspace{-0.13cm}=\hspace{-0.06cm}\mathbf{0} \notag
	\end{align}
for the SRIVC iterations. Recall that the expressions in parentheses in \eqref{conditionlemma2} and in the second sum in \eqref{finiteall} are simply $e(t_k,\bm{\theta}_i^{l+1})$, which follows from the proof of Lemma~\ref{lemmagnsrivc}. Under Assumption \ref{assumption515}, the sums in \eqref{finiteall} converge to their expected values as $N\to\infty$ \cite{soderstrom1975ergodicity}. Thus, as $N\to \infty$,
\begin{equation}
	\mathbb{E}\hspace{-0.07cm}\left\{ \hspace{-0.05cm}\hat{\bm{\varphi}}_\textnormal{f}(\hspace{-0.01cm}t_k\hspace{-0.01cm},\hspace{-0.025cm}\bm{\theta}_{i}^{l\hspace{-0.02cm}+\hspace{-0.02cm}1}\hspace{-0.035cm}) \hat{\bm{\varphi}}_\textnormal{f}^{\hspace{-0.04cm}\top}\hspace{-0.06cm}(\hspace{-0.01cm}t_k\hspace{-0.01cm},\hspace{-0.025cm}\bm{\theta}_{i}^{l\hspace{-0.02cm}+\hspace{-0.02cm}1}\hspace{-0.035cm})\hspace{-0.025cm} \right\}^{\hspace{-0.02cm}-\hspace{-0.01cm}1} \hspace{-0.03cm}\mathbb{E}\hspace{-0.07cm}\left\{\hspace{-0.05cm} \hat{\bm{\varphi}}_\textnormal{f}(\hspace{-0.01cm}t_k\hspace{-0.01cm},\hspace{-0.025cm}\bm{\theta}_{i}^{l\hspace{-0.02cm}+\hspace{-0.02cm}1}\hspace{-0.035cm}) e(\hspace{-0.01cm}t_k\hspace{-0.01cm},\hspace{-0.025cm}\bm{\theta}_i^{l\hspace{-0.02cm}+\hspace{-0.02cm}1}\hspace{-0.035cm})\hspace{-0.026cm} \right\} \hspace{-0.08cm}=\hspace{-0.05cm} \mathbf{0} \notag
\end{equation}
for the GN iterations, and 
\begin{equation}
	\mathbb{E}\hspace{-0.07cm}\left\{ \hspace{-0.05cm}\hat{\bm{\varphi}}_\textnormal{f}(\hspace{-0.01cm}t_k\hspace{-0.01cm},\hspace{-0.025cm}\bm{\theta}_{i}^{l\hspace{-0.02cm}+\hspace{-0.02cm}1}\hspace{-0.035cm}) \bm{\varphi}_\textnormal{f}^{\hspace{-0.04cm}\top}\hspace{-0.06cm}(\hspace{-0.01cm}t_k\hspace{-0.01cm},\hspace{-0.025cm}\bm{\theta}_{i}^{l\hspace{-0.02cm}+\hspace{-0.02cm}1}\hspace{-0.035cm})\hspace{-0.025cm} \right\}^{\hspace{-0.02cm}-\hspace{-0.01cm}1} \hspace{-0.03cm}\mathbb{E}\hspace{-0.07cm}\left\{\hspace{-0.05cm} \hat{\bm{\varphi}}_\textnormal{f}(\hspace{-0.01cm}t_k\hspace{-0.01cm},\hspace{-0.025cm}\bm{\theta}_{i}^{l\hspace{-0.02cm}+\hspace{-0.02cm}1}\hspace{-0.035cm}) e(\hspace{-0.01cm}t_k\hspace{-0.01cm},\hspace{-0.025cm}\bm{\theta}_i^{l\hspace{-0.02cm}+\hspace{-0.02cm}1}\hspace{-0.035cm})\hspace{-0.026cm} \right\} \hspace{-0.08cm}=\hspace{-0.05cm} \mathbf{0} \notag
\end{equation}
for the SRIVC iterations. Either way, under Assumption \ref{assumption53}, the converging point must (generically) satisfy
\begin{equation}
	\label{precondition}
	\mathbb{E}\left\{ \hat{\bm{\varphi}}_\textnormal{f}(t_k,\bm{\theta}_{i}^{l+1}) e(t_k,\bm{\theta}_i^{l+1}) \right\} = \mathbf{0}.
\end{equation}
The instrument vector $\hat{\bm{\varphi}}_\textnormal{f}$ can be written as
\begin{equation}
	\label{hatvarphi}
	\hat{\bm{\varphi}}_\textnormal{f}(t_k,\bm{\theta}_{i}^{l+1}) = \mathcal{S}(-B_i^{l+1},A_i^{l+1}) \frac{1}{[A_i^{l+1}(p)]^2}\mathbf{u}_i(t_k), \hspace{-0.1cm}
\end{equation}
where $B_i^{l+1},A_i^{l+1}$ denote the $i$th model polynomials evaluated at $\bm{\theta}_{i}^{l+1}$, $\mathbf{u}_i(t_k)$ has the form in \eqref{ujt_k}, and the Sylvester matrix $\mathcal{S}(-B_i^{l+1},A_i^{l+1})$ is non-singular thanks to $B_i^{l+1}$ and $A_i^{l+1}$ being coprime \cite[Lemma A3.1]{soderstrom1983instrumental}. On the other hand, $e(t_k,\bm{\theta}_i^{l+1})$ can be written as
\vspace{-0.1cm}
\begin{equation}
	\label{et_k2}
	e(t_k,\bm{\theta}_i^{l+1})= \sum_{j=1}^{K} \bigg[\frac{B_j^*(p)}{A_j^*(p)}-\frac{\bar{B}_j(p)}{\bar{A}_j(p)}\bigg]u_j(t_k) + v(t_k),
\end{equation}
where we have used the notation in \eqref{notationAjBj} for defining $\bar{B}_j(p)/\bar{A}_j(p)$. By exploiting these alternative descriptions for $\hat{\bm{\varphi}}_\textnormal{f}$ and $e$, we can write the condition \eqref{precondition} as	

\vspace{-0.3cm}
\small{\begin{align}
	\sum_{j=1}^K \hspace{-0.04cm}\mathbb{E}\hspace{-0.03cm}\bigg\{\hspace{-0.08cm}\frac{1}{\bar{A}_i^2\hspace{-0.03cm}(\hspace{-0.01cm}p\hspace{-0.01cm})}\mathbf{u}_i\hspace{-0.03cm}(\hspace{-0.01cm}t_k\hspace{-0.01cm}) \hspace{-0.05cm}\bigg[\hspace{-0.03cm}\frac{B_{\hspace{-0.01cm}j}^{\hspace{-0.01cm}*}\hspace{-0.04cm}(\hspace{-0.01cm}p\hspace{-0.01cm})}{A_{\hspace{-0.01cm}j}^{\hspace{-0.01cm}*}\hspace{-0.04cm}(\hspace{-0.01cm}p\hspace{-0.01cm})}\hspace{-0.07cm}-\hspace{-0.07cm}\frac{\bar{B}_{\hspace{-0.02cm}j}\hspace{-0.03cm}(\hspace{-0.01cm}p\hspace{-0.01cm})}{\bar{A}_{\hspace{-0.02cm}j}\hspace{-0.03cm}(\hspace{-0.01cm}p\hspace{-0.01cm})}\hspace{-0.03cm}\bigg]\hspace{-0.03cm}u_{\hspace{-0.01cm}j}\hspace{-0.02cm}(\hspace{-0.01cm}t_k\hspace{-0.01cm}) \hspace{-0.05cm}\bigg\} \hspace{-0.1cm}&=\hspace{-0.07cm}\mathbb{E}\hspace{-0.03cm}\bigg\{\hspace{-0.09cm}\frac{1}{\bar{A}_i^2\hspace{-0.03cm}(\hspace{-0.01cm}p\hspace{-0.01cm})}\mathbf{u}_i\hspace{-0.02cm}(\hspace{-0.013cm}t_k\hspace{-0.013cm}) v\hspace{-0.01cm}(\hspace{-0.013cm}t_k\hspace{-0.013cm})\hspace{-0.05cm}\bigg\} \notag \\
        &=\mathbf{0}, \notag
\end{align}}
\normalsize
\hspace{-0.11cm}where the last equality holds because the signals $u$ and $v$ are uncorrelated due to Assumption \ref{assumption515}. Thus, \eqref{precondition} reduces to
\begin{equation}
	\sum_{j=1}^K\mathbb{E}\bigg\{\frac{1}{\bar{A}_i^2(p)}\mathbf{u}_i(t_k) \bigg[\frac{B_j^*(p)}{A_j^*(p)}-\frac{\bar{B}_j(p)}{\bar{A}_j(p)}\bigg] u_j(t_k)\bigg\} =\mathbf{0}. \notag
\end{equation}
Finally, by rearranging the residual terms as
\begin{equation}
	\bigg[\frac{B_j^*(p)}{A_j^*(p)}-\frac{\bar{B}_j(p)}{\bar{A}_j(p)}\bigg] u_j(t_k) = \frac{1}{A_j^*(p)\bar{A}_j(p)} \mathbf{u}_j^{\top}(t_k)\bm{\eta}_j, \notag 
\end{equation}
the converging point $\bm{\theta}_i^{l\hspace{-0.02cm}+\hspace{-0.02cm}1}$ must satisfy \eqref{conditionadditive}.
\end{proof}
\begin{corollary}
\label{coromiso}
    Under the same assumptions of Theorem~\ref{thmmiso}, for the MISO system case, the converging point of the $i$th model for a fixed descent iteration as $N$ tends to infinity generically satisfies $\bar{B}_i(p)/\bar{A}_i(p) = B_i^*(p)/A_i^*(p)$.
\end{corollary}
\begin{proof}
All expectations in \eqref{conditionadditive} are zero except for when $j=i$, due to the assumption that the inputs $u_i$ are uncorrelated. Thus, the expectation gives a matrix that is generically non-singular due to the genericity argument of Theorem 1 of \cite{pan2020consistency}. We reach $\bm{\eta}_i=\mathbf{0}$, which occurs if and only if $\bar{B}_i(p)/\bar{A}_i(p) = B_i^*(p)/A_i^*(p)$.
\end{proof}

Theorem \ref{thmmiso} reveals that, at a fixed descent iteration $l$, the block coordinate descent algorithm for additive SISO models will lead to biased estimates in general, and more bias is expected if the other submodels deviate further from their true subsystem. 
On the other hand, Corollary \ref{coromiso} reveals that consistent estimates can be obtained in the MISO framework by a single descent iteration with GN or SRIVC iterations performed until convergence. Such approach is similar but not equivalent to the one in \cite{garnier2007optimal}, which performs one SRIVC iteration for each submodel, and descent iterations until convergence. Descent is not guaranteed with only one SRIVC iteration, which suggests that the method in \cite{garnier2007optimal} can benefit from multiple GN or SRIVC iterations for each submodel.

\subsection{Persistence of excitation of the input}

Theorem \ref{thmmiso} requires a persistence of excitation order of $2\max_i n_i$ for the additive SISO case. However, such condition is not enough to avoid identifiability issues when trying to estimate each submodel separately. An example of this identifiability problem is detailed below.

\begin{example}
	\label{example2}
	For the additive SISO case, let $G_1^*(p)$ and $G_2^*(p)$ be two strictly causal, asymptotically stable, first-order subsystems. If the input is a sampled sinusoid signal (with ZOH intersample behavior and sampling period $h$) of the form $u(t_k)=\sin(\omega kh)$, with $|\omega|<\pi/h$, then, as $N\to \infty$, the GN or SRIVC iterations for estimating $G_1^*(p)$ lead to the following interpolation conditions:
	\begin{equation}
		G_{\hspace{-0.02cm}\textnormal{d},\hspace{-0.02cm}1}\hspace{-0.04cm}(\hspace{-0.01cm}e^{\hspace{-0.02cm}\pm\mathrm{i}\omega\hspace{-0.02cm}h}\hspace{-0.05cm},\hspace{-0.03cm}\bm{\theta}_1^{l\hspace{-0.02cm}+\hspace{-0.02cm}1}\hspace{-0.02cm})\hspace{-0.06cm}=\hspace{-0.06cm}G_{\hspace{-0.02cm}\textnormal{d},\hspace{-0.02cm}1}^*\hspace{-0.04cm}(\hspace{-0.02cm}e^{\hspace{-0.02cm}\pm\mathrm{i}\omega \hspace{-0.03cm}h}\hspace{-0.01cm}) +G_{\hspace{-0.02cm}\textnormal{d},\hspace{-0.02cm}2}^*(\hspace{-0.02cm}e^{\hspace{-0.02cm}\pm\mathrm{i}\omega\hspace{-0.03cm}h}\hspace{-0.01cm})-G_{\textnormal{d},\hspace{-0.01cm}2}\hspace{-0.01cm}(\hspace{-0.02cm}e^{\hspace{-0.02cm}\pm\mathrm{i}\omega \hspace{-0.01cm}h}\hspace{-0.03cm},\hspace{-0.02cm}\bm{\theta}_2^{l}), \notag
	\end{equation}
	where $G_{\textnormal{d},i}$ denotes the ZOH discrete-time equivalent of the continuous-time transfer function $G_i$. On the other hand, the GN or SRIVC iterations for estimating $G_2^*(p)$ converge to
	\begin{equation}
		G_{\hspace{-0.02cm}\textnormal{d}\hspace{-0.01cm},\hspace{-0.02cm}2}\hspace{-0.04cm}(\hspace{-0.02cm}e^{\hspace{-0.02cm}\pm\mathrm{i}\omega\hspace{-0.02cm}h}\hspace{-0.05cm},\hspace{-0.03cm}\bm{\theta}_2^{l\hspace{-0.02cm}+\hspace{-0.02cm}1}\hspace{-0.045cm})\hspace{-0.08cm}=\hspace{-0.06cm}G_{\hspace{-0.03cm}\textnormal{d},\hspace{-0.02cm}1}^*\hspace{-0.05cm}(\hspace{-0.04cm}e^{\hspace{-0.02cm}\pm\mathrm{i}\omega \hspace{-0.03cm}h}\hspace{-0.04cm}) \hspace{-0.03cm}+G_{\hspace{-0.04cm}\textnormal{d}\hspace{-0.01cm},\hspace{-0.02cm}2}^*\hspace{-0.03cm}(\hspace{-0.03cm}e^{\hspace{-0.02cm}\pm\mathrm{i}\omega\hspace{-0.03cm}h}\hspace{-0.04cm})-G_{\textnormal{d},\hspace{-0.01cm}1}\hspace{-0.01cm}(\hspace{-0.03cm}e^{\hspace{-0.02cm}\pm\mathrm{i}\omega \hspace{-0.01cm}h}\hspace{-0.03cm},\hspace{-0.02cm}\bm{\theta}_1^{l\hspace{-0.02cm}+\hspace{-0.02cm}1}\hspace{-0.02cm}). \notag
	\end{equation}
	These equations indicate that $G_{\textnormal{d},2}(e^{\pm\mathrm{i}\omega\hspace{-0.02cm} h}\hspace{-0.04cm},\bm{\theta}_2^{l\hspace{-0.02cm}+\hspace{-0.02cm}1})=G_{\textnormal{d},2}(e^{\pm\mathrm{i}\omega\hspace{-0.02cm} h}\hspace{-0.04cm},\bm{\theta}_2^{l})$, i.e., $\bm{\theta}_2$ converges in one descent iteration. Thus, the parameters $\bm{\theta}_1^{l\hspace{-0.02cm}+\hspace{-0.02cm}1}$ and $\bm{\theta}_2^{l\hspace{-0.02cm}+\hspace{-0.02cm}1}$ must satisfy the underdetermined system of equations
	\begin{equation}
		G_{\hspace{-0.02cm}\textnormal{d}\hspace{-0.01cm},\hspace{-0.02cm}1}\hspace{-0.045cm}(\hspace{-0.03cm}e^{\hspace{-0.02cm}\pm\mathrm{i}\omega\hspace{-0.02cm}h}\hspace{-0.05cm},\hspace{-0.03cm}\bm{\theta}_1^{l\hspace{-0.02cm}+\hspace{-0.02cm}1}\hspace{-0.04cm})+G_{\hspace{-0.02cm}\textnormal{d}\hspace{-0.01cm},\hspace{-0.02cm}2}\hspace{-0.03cm}(\hspace{-0.03cm}e^{\hspace{-0.02cm}\pm\mathrm{i}\omega \hspace{-0.03cm}h}\hspace{-0.05cm},\hspace{-0.03cm}\bm{\theta}_2^{l\hspace{-0.02cm}+\hspace{-0.02cm}1}\hspace{-0.02cm}) \hspace{-0.09cm}=\hspace{-0.06cm}G_{\hspace{-0.03cm}\textnormal{d}\hspace{-0.01cm},\hspace{-0.02cm}1}^*\hspace{-0.04cm}(\hspace{-0.03cm}e^{\pm\mathrm{i}\omega \hspace{-0.03cm}h}\hspace{-0.03cm}) +G_{\hspace{-0.03cm}\textnormal{d}\hspace{-0.01cm},\hspace{-0.02cm}2}^*(\hspace{-0.03cm}e^{\hspace{-0.02cm}\pm\mathrm{i}\omega\hspace{-0.03cm}h}\hspace{-0.03cm}), \notag 
	\end{equation}
	which is the expected result for identification with a single sinusoidal input. Since an infinite number of solutions exist for the system above, the parameters are not identifiable.  \hfill $\triangle$ 
\end{example} 
\begin{remark}
	A similar observation to Example \ref{example2} can be derived for \textit{continuous-time} multisine inputs. In such case, for $u(t) = \sin(\omega_1 t), t\geq 0$, a tailored version of the SRIVC iterations suited for continuous-time multisines \cite{gonzalez2020consistent} results in $G_1\hspace{-0.01cm}(\pm\mathrm{i}\omega_1,\bm{\theta}_1^{l+1}\hspace{-0.01cm})\hspace{-0.03cm}+\hspace{-0.03cm}G_2\hspace{-0.03cm}(\hspace{-0.01cm}\pm\hspace{-0.01cm}\mathrm{i}\omega_1\hspace{-0.01cm},\bm{\theta}_2^{l\hspace{-0.01cm}+\hspace{-0.01cm}1}\hspace{-0.01cm})\hspace{-0.03cm} =\hspace{-0.03cm}G_1^*\hspace{-0.01cm}(\hspace{-0.01cm}\pm\hspace{-0.01cm}\mathrm{i}\omega_1\hspace{-0.01cm}) \hspace{-0.03cm}+\hspace{-0.03cm}G_2^*\hspace{-0.01cm}(\hspace{-0.01cm}\pm\hspace{-0.01cm}\mathrm{i}\omega_1\hspace{-0.01cm})$, which again indicates a problem of identifiability.
 \end{remark}	
 
Towards the goal of solving this identifiability problem, the next subsection studies sufficient conditions for consistency of the coordinate descent method for both MISO and additive SISO identification.
\vspace{-0.1cm}

\subsection{Consistency analysis: infinite descent iterations}
\label{subrevisited}
We first must investigate the nonsingularity of a certain matrix formed by filtered versions of the input signal.

\begin{lemma}
	\label{lemmaphi}
	Let the input $u(t_k)$ be persistently exciting of order no less than $2\sum_{i=1}^K n_i$, and assume that the sampling frequency is more than twice that of the largest imaginary part of the zeros of $\prod_{j=1}^K \bar{A}_j(p)A_j^*(p)$. Moreover, suppose Assumptions \ref{assumption51} to \ref{assumption52} hold. Then, the matrix 
	\begin{equation}
		\label{phimatrix}
		\bm{\Phi} = \mathbb{E}\hspace{-0.03cm}\left\{\hspace{-0.02cm}
		\begin{bmatrix}
			\frac{1}{\bar{A}_1^2\hspace{-0.02cm}(p)}\mathbf{u}_1\hspace{-0.02cm}(t_k) \\
			\frac{1}{\bar{A}_2^2\hspace{-0.02cm}(p)}\mathbf{u}_2\hspace{-0.02cm}(t_k) \\
			\vdots \\
			\frac{1}{\bar{A}_K^2\hspace{-0.02cm}(p)}\mathbf{u}_K\hspace{-0.02cm}(t_k)
		\end{bmatrix}
		\begin{bmatrix}
			\frac{1}{A_1^*(p)\bar{A}_1(p)}\mathbf{u}_1\hspace{-0.02cm}(t_k) \\
			\frac{1}{A_2^*(p)\bar{A}_2(p)}\mathbf{u}_2\hspace{-0.02cm}(t_k) \\
			\vdots \\
			\frac{1}{A_K^*(p)\bar{A}_K(p)}\mathbf{u}_K\hspace{-0.02cm}(t_k)
		\end{bmatrix}^\top \right\},
	\end{equation}
	with $\mathbf{u}_j(t_k)$ defined in \eqref{ujt_k}, is generically non-singular with respect to the system and model denominator polynomials.
\end{lemma}
\begin{proof}
Direct from Lemma 3 of \cite{gonzalez2023identification} when, following their notation, we set $S(q)=1$ and $z(t_k)=u(t_k)$.    
\end{proof}
Theorem \ref{thmconsistencyfinal} shows that the block coordinate descent method possesses the consistency property in both MISO and additive SISO setups, provided the persistence of excitation order of the input in the additive SISO case is at least twice the number of poles of the aggregated model. 
\begin{theorem}
\label{thmconsistencyfinal}
	Consider Algorithm \ref{algorithm1}, and suppose that Assumptions \ref{assumption51} to \ref{assumption52} hold. In addition, assume that the input $u(t_k)$ is persistently exciting of order no less than $2\sum_{i=1}^K n_i$ for the additive SISO setup, and that the sampling frequency is more than twice that of the largest imaginary part of the zeros of $\prod_{j=1}^K \bar{A}_j(p)A_j^*(p)$. If the GN or SRIVC iterations are performed until convergence, and the descent iterations $l\to\infty$, then the true parameters $\{\bm{\theta}_i^*\}_{i=1}^K$ are the unique converging point (generically) as $N\to \infty$.
\end{theorem}
\begin{proof}
As the number of GN or SRIVC iterations tends to infinity, the model parameters satisfy \eqref{conditionadditive}. At convergence in descent iterations, the condition $\bm{\Phi}\bm{\eta}=\mathbf{0}$ is obtained, where $\bm{\Phi}$ is given by \eqref{phimatrix} and $\bm{\eta}:= [\bm{\eta}_1^\top, \dots,\bm{\eta}_K^\top]^\top$ is the joint bias vector. Lemma \ref{lemmaphi} shows that $\bm{\Phi}$ is generically non-singular, which implies that $\bm{\eta}=\mathbf{0}$ generically. This expression means that $\bar{B}_i(p)/\bar{A}_i(p) = B_i^*(p)/A_i^*(p)$ for all $i=1,\dots,K$, which proves the statement.    
\end{proof}
\begin{remark}
    There exists a parameter identifiability issue in the additive SISO setup when a pair of subsystems shares the same model structure. Swapping such pair results in an identical additive model, altering the arrangement of parameters within $\bm{\beta}$. This property, however, does not impact the consistency of the additive model estimator as a whole.
    \end{remark}

\vspace{-0.2cm}
\section{Simulations}
\label{sec:simulations}
In this section, we will verify the theoretical results via Monte Carlo simulations. We consider the following systems:
\begin{equation}
    G_1^*\hspace{-0.01cm}(p) \hspace{-0.05cm} = \hspace{-0.05cm}\frac{2}{0.25p^2\hspace{-0.03cm}+\hspace{-0.03cm}0.25p\hspace{-0.03cm}+\hspace{-0.05cm}1}, \quad \hspace{-0.14cm} G_2^*(p)\hspace{-0.05cm}=\hspace{-0.05cm}\frac{1}{0.025p^2\hspace{-0.03cm}+\hspace{-0.03cm}0.01p\hspace{-0.03cm}+\hspace{-0.05cm}1}. \notag
\end{equation}
The output is contaminated by white noise of variance $0.25$, and is sampled every $0.02[\textnormal{s}]$. Each setup considers unit-variance Gaussian white noise inputs, both resulting in an output signal-to-noise ratio of $6.6$ [dB]. The additive SISO and MISO estimators consider SRIVC iterations; GN iterations have shown no significant performance differences at convergence in iterations, in agreement with Lemma \ref{lemmagnsrivc}.

\subsection{Bias at a fixed descent iteration}
\label{subsec:biasexperiment}
\begin{figure}
	\centering{
		\includegraphics[width=0.475\textwidth]{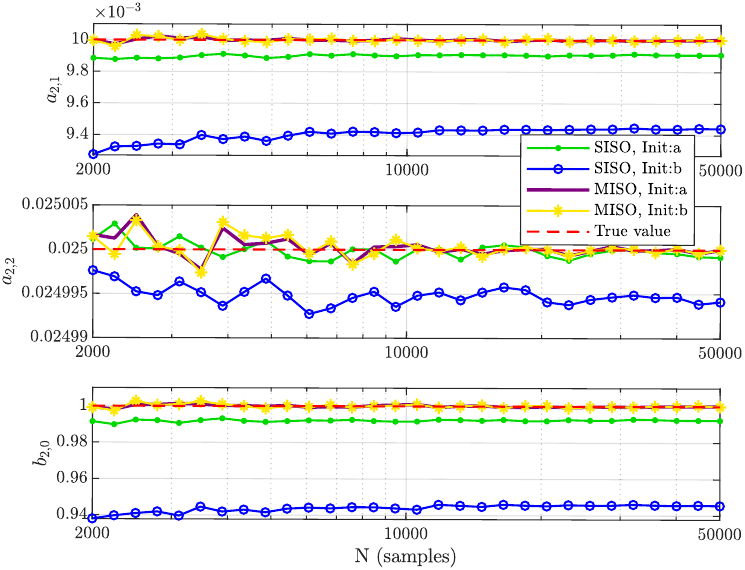}
  \vspace{-0.3cm}
		\caption{Empirical mean of the estimate of each parameter of $G_2^*(p)$ for one descent iteration. The true values are given by $a_{2,1}^*=0.01$, $a_{2,2}^*=0.025$, and $b_{2,0}^*=1$. The estimates of the MISO model setup converge to the true value, while bias is observed in for the additive SISO setup.}
		\label{fig1}}
\end{figure}

To verify the results in Theorem \ref{thmmiso}, we perform 300 Monte Carlo runs with varying input and noise realizations for 30 different sample sizes, ranging logarithmically from $N=2000$ to $N=50000$. On each run, one descent iteration is performed on $G_2(p)$ for a fixed model estimate $\bar{G}_1(p) = \bar{B}_1(p)/\bar{A}_1(p)$. The additive SISO and MISO setups are tested under two different model estimates for $\bar{G}_1(p)$: 
\begin{equation}
    \bar{G}_1^{\textnormal{a}}\hspace{-0.01cm}(p) \hspace{-0.05cm}  = \hspace{-0.05cm}\frac{2.2}{0.2p^2\hspace{-0.03cm}+\hspace{-0.03cm}0.2p\hspace{-0.03cm}+\hspace{-0.05cm}1}, \quad \hspace{-0.14cm} \bar{G}_1^\textnormal{b}(p)\hspace{-0.05cm}=\hspace{-0.05cm}\frac{1.7}{0.1p^2\hspace{-0.03cm}+\hspace{-0.03cm}0.27p\hspace{-0.03cm}+\hspace{-0.05cm}1}. \notag
\end{equation}
The 2-norm errors of each initial model for $G_1^*(p)$ are $\|\bar{G}_1^{\textnormal{a}}-G_1^*\|_2 = 1.25$ and $\|\bar{G}_1^{\textnormal{b}}-G_1^*\|_2 = 2.11$. The empirical means of each parameter estimate of $G_2^*(p)$ are shown in Figure \ref{fig1}. As expected from Corollary \ref{coromiso}, the MISO estimates converge to the true values for increasing sample size, independently of the initial estimate of $G_1^*(p)$. In contrast, the coordinate descent step for the additive SISO setup suffers from bias in the parameter estimates. Since $\bar{G}_1^\textnormal{b}(p)$ is a more inaccurate estimate of $G_1^*(p)$ compared to $\bar{G}_1^\textnormal{a}(p)$, its associated bias vector $\bm{\eta}_1$ in Theorem \ref{thmmiso} is larger in magnitude, implying that $\bm{\eta}_2$ must also be large for \eqref{conditionadditive} to be satisfied. The sample means for $G_2(p)$ are further away from their true values if $\bar{G}_1^\textnormal{b}(p)$ is used as an estimate of $G_1^*(p)$, which is in accordance to this insight.

\vspace{-0.3cm}
\subsection{Consistency}
  \vspace{-0.1cm}
The consistency of the block coordinate descent estimator for additive SISO models is studied under a similar experimental condition to Section \ref{subsec:biasexperiment}. This time, the estimators are both initialized with $\bar{G}_1^{\textnormal{b}}(p)$, and ten SRIVC iterations are performed for each descent step. A maximum of 30 descent iterations are used for each estimate, and the iterations also may terminate if the relative error of the parameter vector with respect to its previous value is below $10^{-10}$. The estimated parameters for $G^*_2(p)$ are analyzed. To avoid exchangeability issues for the additive SISO setup, we study the submodel parameters that are closest to $G_2^*(p)$ according to the 2-norm of the parameter error.

\begin{figure}
\vspace{-0.2cm}
	\centering{
		\includegraphics[width=0.475\textwidth]{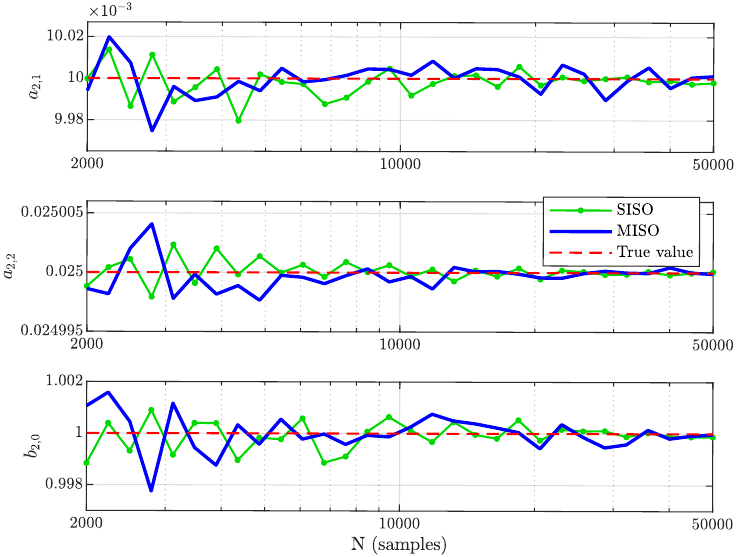}
  \vspace{-0.3cm}
		\caption{Empirical mean of the estimate of each parameter of $G_2^*(p)$ using the full block coordinate descent method. The true values are given by $a_{2,1}^*=0.01$, $a_{2,2}^*=0.025$, and $b_{2,0}^*=1$. Both the MISO and additive SISO estimators converge to the true values as the sample size grows.}
		\label{fig2}}
\vspace{-0.6cm}
\end{figure}
The results in Fig. \ref{fig2} show that both MISO and additive SISO setups achieve consistent estimates of $G_2^*(p)$. This result is in accordance to Theorem \ref{thmconsistencyfinal}. Note that the persistence of excitation is not an issue in this test, since the inputs are designed to be persistently exciting of any order.

\section{Conclusions}
\label{sec:conclusions}
\vspace{-0.02cm}
In this paper we have analyzed the statistical properties of the block coordinate descent method applied to the identification of MISO and additive SISO systems in open loop. The main difference between these setups is the nature of the input excitation, which reveals different bias properties and conditions for consistency. We have characterized the bias after a single descent for additive SISO systems, and we have used this insight to prove the generic consistency of the method for MISO and additive SISO setups in a unified manner. The results hold for both Gauss-Newton and SRIVC iterations. The numerical experiments here presented have verified the theoretical results.
\bibliography{References}
\end{document}